\documentclass[aps,prl,reprint,superscriptaddress,preprintnumbers,amsmath,amssymb]{revtex4-2}

\usepackage{graphicx}
\usepackage{bm}
\usepackage[utf8]{inputenc}
\usepackage{color}
\usepackage[colorlinks=true, letterpaper=true, pdfstartview=FitV, linkcolor=blue, citecolor=blue, urlcolor=blue]{hyperref}

\newcommand{\CV}[0]{\color{blue}}
\newcommand{\CIV}[0]{\color{black}}

\begin{document}

\title{Correlated \texorpdfstring{$\mathcal{PT}$}{PT}-Symmetric Antiferromagnetic Topological Insulators with Giant Nonlinear Anomalous Thermoelectrics}

\author{Heng-Yu Di}
\affiliation{School of Physical Sciences, University of Chinese Academy of Sciences, Beijing 100049, China}

\author{Zhen-Gang Zhu}
\email{zgzhu@ucas.ac.cn}
\affiliation{School of Physical Sciences, University of Chinese Academy of Sciences, Beijing 100049, China}
\affiliation{School of Electronic, Electrical and Communication Engineering, University of Chinese Academy of Sciences, Beijing 100049, China}

\author{Gang Su}
\email{gsu@ucas.ac.cn}
\affiliation{Institute of Theoretical Physics, Chinese Academy of Sciences, Beijing 100190, China}
\affiliation{School of Physical Sciences, University of Chinese Academy of Sciences, Beijing 100049, China}
\affiliation{Kavli Institute for Theoretical Sciences, University of Chinese Academy of Sciences, Beijing 100190, China}

\date{\today}

\begin{abstract}
Topological states in antiferromagnets (AFMs) offer a promising platform for exploring novel physical phenomena and advancing the applications of AFM spintronics. The AFM topological insulator (TI) state stands out as one of the most representative and prominent cases. Unlike the previously proposed AFM-TI states in noninteracting systems, here we employ an extended Kane-Mele-Hubbard model to demonstrate that electron correlations can give rise to a $\mathcal{PT}$-symmetric AFM-TI state. This state breaks both spatial inversion symmetry $\mathcal{P}$ and time-reversal symmetry $\mathcal{T}$, and enables intrinsic topological nonlinear responses to dominate the leading-order dynamics of the system. The competition between electron correlations and spin-orbit coupling drives the system across a topological phase transition, where the closure of the bulk band gap induces singular behaviors in higher-order quantum geometric tensors. Such microscopic singular characteristics manifest macroscopically as pronounced enhancements in thermoelectric performance, charge conductivity, and thermal conductivity. These giant tunable transport signatures, which can be effectively modulated by mechanical strain and electrostatic gating, provide a feasible experimental route to probe and understand correlated topological materials.
\end{abstract}

\maketitle

\CV \textit{Introduction.-} \CIV
Introducing ferromagnetic order into topological insulators breaks time-reversal ($\mathcal{T}$) symmetry, yielding exotic states such as the quantum anomalous Hall effect and magnetic Weyl semimetals \cite{Chang2013,Wan2011}.
Antiferromagnetism (AFM), on the other hand, as a distinct aspect of magnetic order that also breaks $\mathcal{T}$ symmetry, has recently drawn significant attention in the field of AFM spintronics \cite{Jungwirth2016}.
Particularly, the three-dimensional (3D) antiferromagnetic topological insulator (AFM-TI) was proposed with $\mathcal{T}$ broken, but $T_{1/2}\mathcal{T}$ symmetry preserved ($T_{1/2}$ is the operator of half-lattice translation), giving rise to the $\mathbb{Z}_{2}$ classification of insulating AFM phases \cite{Mong2010,Fang2013}.
Such an AFM-TI was subsequently observed experimentally in 3D MnBi$_{2}$Te$_{4}$ \cite{Gong2019,Otrokov2019}.
Moreover, a two-dimensional (2D) AFM-TI ensured by a combined symmetry of twofold rotation and half-lattice translation was proposed \cite{Niu2020} but has not been observed yet.

The aforementioned AFM-TIs originate from the band topology of non-interacting systems. In this Letter, we propose a correlation-induced two-dimensional (2D) $\mathcal{PT}$-symmetric AFM-TI phase (where $\mathcal{P}$ denotes inversion symmetry) from a many-body perspective. 
We aim to not only identify this non-trivial topological state but also reveal whether electron correlations can boost topological thermoelectric responses in such systems, with a particular focus on the anomalous Nernst effect (ANE).
Thermoelectric power generation (TPG) based on the ANE exhibits inherent advantages over conventional Seebeck-effect-based TPG \cite{Jian2023,Sakuraba2013,Sakuraba2016,Mizuguchi2019}.
For practical device applications, however, the ANE coefficient requires substantial enhancement.
Although typical topological materials have achieved an ANE coefficient one order of magnitude higher than those of Fe, Co, and Ni-based systems (e.g., a value of approximately $5$ $\mu$V/K in Co$_3$Sn$_2$S$_2$ \cite{Liu2019a,Morali2019,Yang2020,RoyKarmakar2022}), such a magnitude remains insufficient for practical use.
Recently, an ultra-large effective ANE coefficient of nearly $300$ $\mu$V/K was reported in nonmagnetic ABA trilayer graphene, converted from its nonlinear ANE coefficient \cite{Liu2025}.
This prominent response emerges near the charge neutrality point, where strong electron correlations are anticipated.
This inspires our speculation that strong electron correlations may offer a viable route to achieving ANE magnitudes suitable for practical applications.
Furthermore, a giant ANE signal of around $10$ $\mu$V/K has recently been observed in heavy-fermion ferromagnets \cite{Guan2026,Li2026}, where strong electron correlations are regarded as the dominant origin of this phenomenon.
%

\begin{figure*}[t]
\centering
\includegraphics[width=\columnwidth]{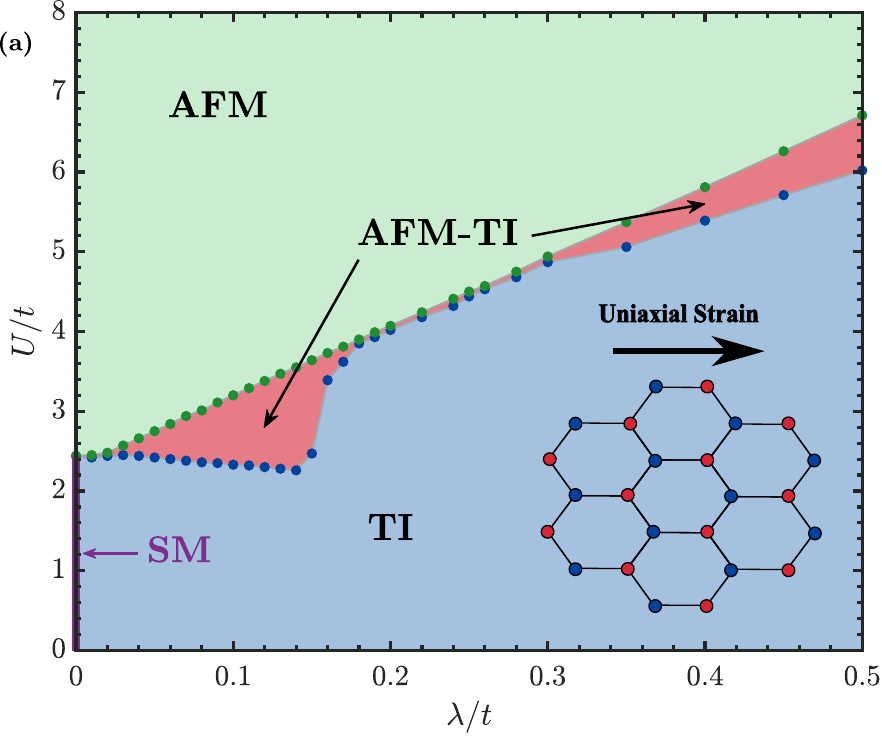}
\hfill
\includegraphics[width=\columnwidth]{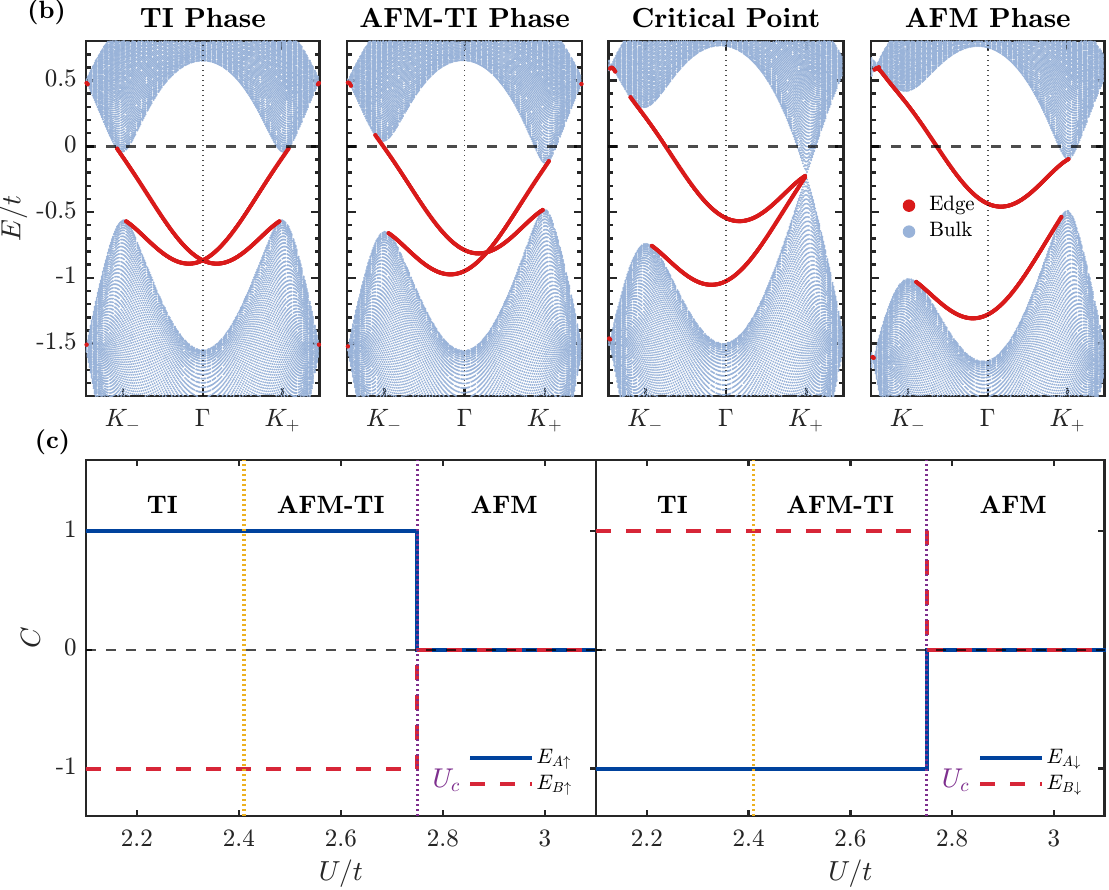}
\caption{\label{fig:phase}
\textbf{Phase diagram and topological transition in the extended Kane-Mele-Hubbard model.}
\textbf{(a)} Ground-state phase diagram in the ($U, \lambda$) parameter space. Parameters are set as: $\eta_1 = 1.05$, $n_A = n_B = 1.005$, $t_2 = 0.1t$, and $k_B T = 0.01t$. Four phases are identified: Semimetal (SM), Topological Insulator (TI), Antiferromagnetic TI (AFM-TI), and trivial Antiferromagnetic Insulator (AFM). The magnetic phase transition from TI to AFM-TI is marked by the onset of a nonzero local magnetization ($m \neq 0$), while the topological transition from AFM-TI phase to AFM phase is determined by the bulk gap closing ($\Delta = 0$). \textit{Inset:} Honeycomb lattice schematic with modified bond vectors $\boldsymbol{\delta}_i$.
\textbf{(b)} Band structures across the four distinct regimes (TI, AFM-TI, critical point, and AFM) at a fixed $\lambda=0.05t$.
\textbf{(c)} Evolution of the band Chern number $C$.
}
\end{figure*}

Guided by the above motivations, we adopt an extended Kane–Mele–Hubbard (KMH) model defined on a honeycomb lattice \cite{Kane2005,Hubbard1963,Rachel2010,Hohenadler2011,Lessnich2024}. We demonstrate that enhancing the on-site Coulomb interaction $U$ stabilizes a 2D $\mathcal{PT}$-symmetric AFM-TI phase, which activates intrinsic nonlinear responses including the ANE, anomalous Hall effect (AHE), and anomalous thermal Hall effect (ATHE). 
Further increasing $U$ drives the system toward a topological phase transition.
At the corresponding critical point, we observe a substantial enhancement of the nonlinear ANE, AHE, and ATHE signals.
Microscopically, such prominent transport enhancement arises from the competition between electron correlations and intrinsic spin–orbit coupling, which induces the divergence of topological quantities near the Dirac point. These pronounced transport signatures serve as robust macroscopic evidence for the realized 2D $\mathcal{PT}$-symmetric AFM-TI phase.
Moreover, we verify that the topological phase transition and associated transport behaviors can be efficiently modulated via mechanical strain and electrostatic gating, offering feasible strategies for future experimental implementation.

\CV \textit{Model and Method.-} \CIV
The KMH model is a prototypical framework for studying the interplay between topology and electron correlations in 2D systems \cite{Rachel2010,Hohenadler2011,Lessnich2024}. To induce the intrinsic nonlinear responses, we investigate an extended KMH model that incorporates next-nearest-neighbor (NNN) hoppings and uniaxial strain. The Hamiltonian reads
\begin{equation}
\begin{aligned}
H  &= \sum_{\langle i,j\rangle}\sum_{\sigma=\uparrow,\downarrow}t_{ij}\left(c^\dagger_{iA\sigma}c_{jB\sigma}+\mathrm{H.c.}\right) \\
   &+ \sum_{\langle\langle i,j\rangle\rangle}\sum_{\gamma=A,B}\sum_{\sigma=\uparrow,\downarrow}t^{\prime}_{ij}\left(c^\dagger_{i\gamma\sigma}c_{j\gamma\sigma}+\mathrm{H.c.}\right) \\
   &+ i\lambda\sum_{\langle\langle i,j\rangle\rangle}\sum_{\gamma=A,B}\sum_{\sigma=\uparrow,\downarrow}\nu_{ij}c^\dagger_{i\gamma\sigma}\tau^{z}_{\sigma\sigma}c_{j\gamma\sigma} \\
   &+ U\sum_{i}\sum_{\gamma=A,B}n_{i\gamma\uparrow}n_{i\gamma\downarrow}.
\label{Hamiltonian}
\end{aligned}
\end{equation}
Here, $c^\dagger_{i\gamma\sigma}$ is the electron creation operator at unit cell $i$ on sublattice $\gamma \in \{A,B\}$ with spin $\sigma \in \{\uparrow, \downarrow\}$. The indices $\langle i,j \rangle$ and $\langle\langle i,j \rangle\rangle$ denote pairs of nearest-neighbor (NN) and NNN sites, respectively.
$t_{ij}$ and $t^{\prime}_{ij}$ denote strain-modified NN and NNN hopping amplitudes. The third term describes the intrinsic Kane-Mele SOC with strength $\lambda$.
$\nu_{ij} = \frac{2}{\sqrt{3}} (\hat{\boldsymbol{d}}_{kj} \times \hat{\boldsymbol{d}}_{ik})_z = \pm 1$ is the Haldane phase factor determined by the orientation of two NN bonds connecting site $j$ to $i$ via the intermediate site $k$. $\tau^z$ is the Pauli matrix acting in spin space.
The last term denotes the on-site Hubbard interaction strength, where $n_{i\gamma\sigma} = c^\dagger_{i\gamma\sigma} c_{i\gamma\sigma}$ is the electron density operator.

In 2D $\mathcal{PT}$-symmetric AFM-TIs, the linear anomalous Hall, Nernst, and thermal conductivities vanish identically \cite{Nagaosa2010,Xiao2010,Xiao2006}, rendering them ineffective for probing the topological state. However, $\mathcal{PT}$ symmetry allows for scattering-independent intrinsic nonlinear topological responses driven by the Berry connection polarizability (BCP) and thermal Berry connection polarizability (TBCP) \cite{Shao2020,Wang2021,Liu2021,Zhang2021a,Wang2022b,Li2024}, establishing them as the primary macroscopic probes for the 2D $\mathcal{PT}$-symmetric AFM-TIs and vice versa.
Characterized by third-rank tensors (e.g., $J_{a} = \sigma^{abc} E_{b} E_{c}$), the nonlinear anomalous electrical ($\sigma^{abc}$), thermoelectric ($\alpha^{abc}$), and thermal ($\kappa^{abc}$) conductivities can be formulated as a unified response tensor $\mathcal{R}_{\zeta}^{abc}$ (see Supplemental Material \cite{SM}):
\begin{equation}
    \mathcal{R}_{\zeta}^{abc} = p_{\zeta} \frac{e^{\frac{(\zeta-2)(\zeta-3)}{2}} k_B^{\zeta}}{\hbar T^{\zeta(2-\zeta)}} \sum_{n} \int [d\boldsymbol{k}] \tilde{\Lambda}_{n}^{abc} F_{\zeta}(f^{\text{FD}}_{n}(\boldsymbol{k}))
\end{equation}
Here, $[d\boldsymbol{k}] = d^2k/(2\pi)^2$, and the integration measure incorporates the weighting function $F_{\zeta}(f^{\text{FD}}) = \int_{0}^{f^{\text{FD}}} [\ln(u^{-1}-1)]^\zeta du$, where $f^{\text{FD}}_n(\boldsymbol{k})$ is the Fermi-Dirac distribution for the $n$-th band. The index $\zeta \in \{0, 1, 2\}$ corresponds to the conductivities $\sigma^{abc}$, $\alpha^{abc}$, and $\kappa^{abc}$. The prefactor is $p_0 = -1$ for charge transport and $p_{1,2} = +1$ for thermal transport. The tensor $\tilde{\Lambda}_{n}^{abc}$ represents the band-resolved BCP when $\zeta=0$, and TBCP when $\zeta=1, 2$.

Uniaxial strain breaks the $C_{3v}$ rotational symmetry of the hexagonal lattice, which is geometrically required to activate the nonlinear topological responses \cite{Yu2019,Liu2021,Zhang2023}.
We parameterize the strain-induced spatial anisotropy of the hoppings through a single independent tuning parameter $\eta_1$. This microscopic hopping ratio maps linearly to the macroscopic uniaxial strain $\epsilon$ (see \cite{SM}).

\CV \textit{Correlation-driven topological phase transitions.-} \CIV
We treat the interaction term within the Hartree-Fock mean-field approximation and determine the local magnetization $m$ and chemical potential $\mu$ for each sublattice self-consistently by minimizing the free energy.
As detailed in the Supplemental Material \cite{SM}, the magnetic phase boundaries obtained for the standard KMH model at half filling without strain quantitatively agree with established mean-field benchmarks \cite{Rachel2010} and qualitatively capture the phase diagram derived from quantum Monte Carlo simulations \cite{Hohenadler2011,Lessnich2024}.
In our extended KMH model study,
we uncover a previously unknown phase, termed $\mathcal{PT}$-symmetric AFM-TI phase, which exists in two prominent regions, one at large $U$ and strong SOC and the other at moderate $U$ and weak SOC [\hyperref[fig:phase]{Fig.~\ref*{fig:phase}(a)}].
Moreover, an unexplored topological phase transition inside the magnetic regime from $\mathcal{PT}$-symmetric AFM-TI to AFM is identified and plays a vital role in the enhancement of topological transport. 

\hyperref[fig:phase]{Fig.~\ref*{fig:phase}(b)} illustrates the band structure evolution of a zigzag ribbon across the phases of TI, AFM-TI, AFM and the critical point.
Robust edge bands crossing the bulk gap and connecting the conduction and valence bands can be seen in both the TI and AFM-TI phases.
The crossing point of these edge bands resides at the $\Gamma$ point in the TI phase due to time reversal symmetry.
However, the crossing point moves away from the $\Gamma$ point to the $K_{+}$ point in AFM-TI phase, reflecting the breaking of $\mathcal{T}$ and $\mathcal{P}$ symmetry in this case.
It is necessary to emphasize that the edge bands are spin-degenerate, so as not to explicitly indicate their spin dependence.
At the critical point, the gap located at $K_{+}$ (shown in \hyperref[fig:phase]{Fig.~\ref*{fig:phase}(b)})and the crossing point of these edge bands move to the $K_{+}$ as well.
Further moving to the AFM phase, it is clearly seen that the edge bands do not cross the gap but only shrink to the conduction or valence band, respectively. Thus, the AFM phase is a topologically trivial phase.
The valley gap can be analytically expressed as $\Delta_{\mathbf{K}} = | \chi(\eta_1)\lambda - m U |$, where $\chi(\eta_1)$ represents the strain-modulated anisotropy factor. The analytical critical condition for the gap closing, $U_c m(U_c) = \chi(\eta_1)\lambda$, agrees with the topological phase transition boundary obtained numerically in \hyperref[fig:phase]{Fig.~\ref*{fig:phase}(a)}, yielding a critical value $U_c = 2.757t$ at $\lambda=0.05t$. The topological nature of this transition is further captured by the band Chern number $C$ [\hyperref[fig:phase]{Fig.~\ref*{fig:phase}(c)}]:
\begin{equation}
  C = \frac{1}{2} \left[ \text{sgn}(\chi(\eta_1)\lambda - mU) + \text{sgn}(\chi(\eta_1)\lambda + mU) \right].
\end{equation}
With increasing correlation $U$, the system undergoes a topological transition from $|C|=1$ to $C=0$ at $U_{c}$.

\begin{figure}[t]
\centering
\includegraphics[width=\columnwidth]{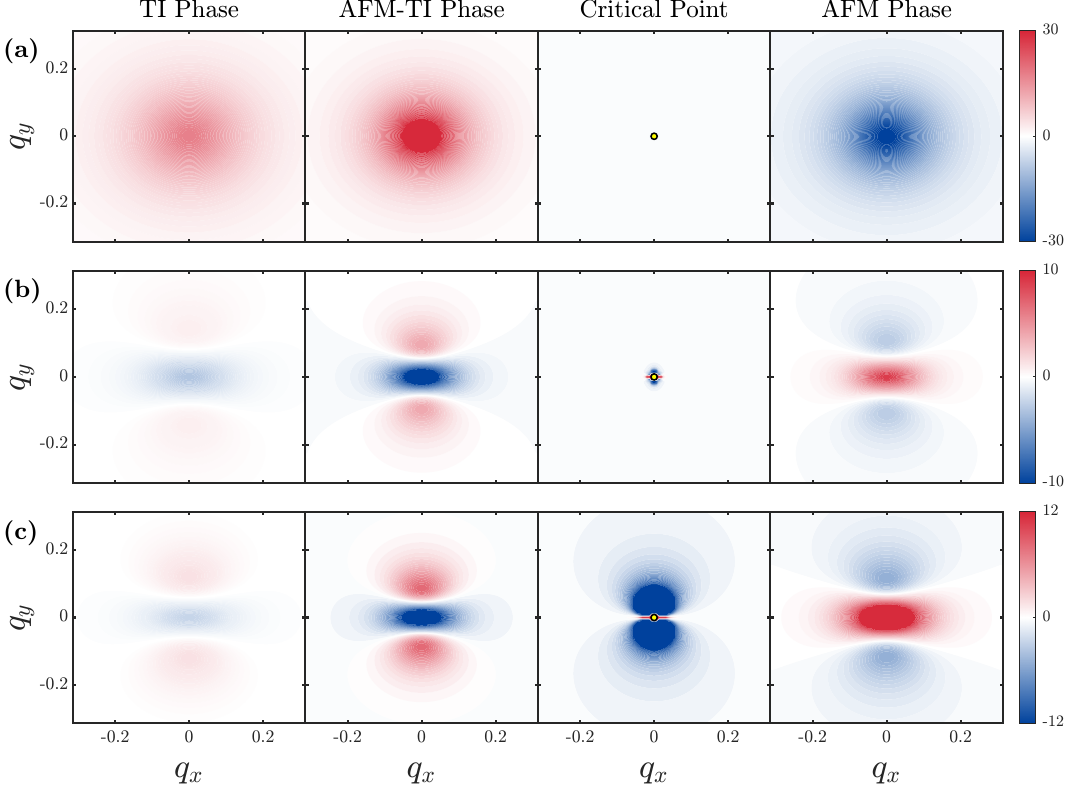}
\caption{\label{fig:topology}
\textbf{Microscopic origin and momentum-space geometric evolution.}
Distributions of \textbf{(a)} the Berry curvature ($\Omega^{yx}$), \textbf{(b)} the $q_y$-even components of the BCP dipole ($\Lambda^{yxx}_{\mathrm{even}}$), and \textbf{(c)} the TBCP dipole ($\Lambda^{yxx,t}_{\mathrm{even}}$) near the Dirac valley $\mathbf{K}_{+}$. The four columns illustrate the evolution across four distinct regimes at a fixed $\lambda=0.05t$. The yellow dot in the third column marks the gapless Dirac point where the geometric quantities diverge.
}
\end{figure}

\CV \textit{Quantum geometric evolution.-} \CIV
While the global Chern number captures the distinct topological phases, the nonlinear anomalous transport is governed by the local momentum-space geometry. We derive the low-energy effective Hamiltonian near the Dirac points ($\mathbf{K}_{\pm}$). The analytical expressions for the Berry curvature ($\Omega$), the BCP dipole ($\Lambda$), and the TBCP dipole ($\Lambda^t$) for a given valley $w_{\xi}=\pm 1$, spin $s_{\sigma}=\pm 1$, and sublattice $\nu_{\gamma}=\pm 1$ read
\begin{eqnarray}
\Omega^{yx}_{\gamma\sigma\xi} &=& -\nu_{\gamma}s_{\sigma}w_{\xi} \frac{m_{\xi,1}t^2\tan\varphi}{2\sqrt{3}(m^2_{\xi}+t^2q^{\prime 2})^{\frac{3}{2}}},
\label{Omegayx} \\
\Lambda_{\gamma\sigma\xi}^{yxx} &=& \frac{\nu_{\gamma}s_{\sigma}t^2}{2(m_{\xi}^2+t^2q^{\prime 2})^{\frac{5}{2}}} \left(m_{\xi}m_{\xi,2} + \frac{\tan^2\varphi}{3}t^2q_{y}\right),
\label{Lambdayxx} \\
%
\Lambda_{\gamma\sigma\xi}^{yxx,t} &=&\frac{\nu_{\gamma}s_{\sigma}t^2}{2(m_{\xi}^2+t^2q^{\prime 2})^{\frac{5}{2}}} \left[ n_{\xi} \left(m_{\xi}m_{\xi,2} + \frac{1}{3}t^2\tan^2\varphi q_{y}\right)\right. \notag\\
&  -& \left. \frac{n_{\xi,2}}{2}\left(m^2_{\xi} + \frac{1}{3}t^2\tan^2\varphi q^2_{y}\right) \right].
\label{Lambdayxxt}
\end{eqnarray}
Here, $q^{\prime} = \sqrt{q_x^2 + \frac{1}{3}\tan^2\varphi q_{y}^2}$ denotes the scaled momentum, and the structural anisotropy is defined by $\cos\varphi = 1/(2\eta_1)$. The critical scaling behavior is governed by $m_{\xi} = m_{\xi,1} + m_{\xi,2}q_{y}$ and $n_{\xi} = n_{\xi,1} + n_{\xi,2}q_{y}$, where the parameter $m_{\xi,1}$ captures the competition between the correlation-induced antiferromagnetic order and the intrinsic SOC (see \cite{SM}).

Based on the effective model, we numerically evaluate the momentum-space distributions of these quantum geometric quantities, utilizing the same parameters as in \hyperref[fig:phase]{Fig.~\ref*{fig:phase}(a)}.
\hyperref[fig:topology]{Fig.~\ref*{fig:topology}} illustrates the evolution of the Berry curvature, alongside the $q_y$-even components of the BCP and TBCP dipoles at the $\mathbf{K}_{+}$ valley.
We explicitly focus on these $q_y$-even components as they dominate the full Brillouin zone transport integrals governed by even-parity Fermi weighting functions. 
In the TI phase, intact $\mathcal{P}$ symmetry and $\mathcal{T}$ symmetry dictate a finite linear spin anomalous transport but vanishing nonlinear anomalous transport, as the BCP and TBCP dipoles exhibit opposite signs at opposite valleys (see \cite{SM}).
Entering the AFM-TI phase, the topology-driven breaking of $\mathcal{P}$ and $\mathcal{T}$ symmetries lifts this valley degeneracy, activating a net nonlinear contribution. 
As $U$ reaches the topological critical point $U_c$, the energy gap at $\mathbf{K}_{+}$ completely closes, resulting in a divergence of these geometric quantities exactly at the gapless Dirac point. 
In the AFM phase, the fundamental topological transition directly flips the signs of the geometric quantities at $\mathbf{K}_{+}$ and rapidly suppresses their magnitudes as the trivial gap reopens.

\begin{figure}[t]
\centering
\includegraphics[width=\columnwidth]{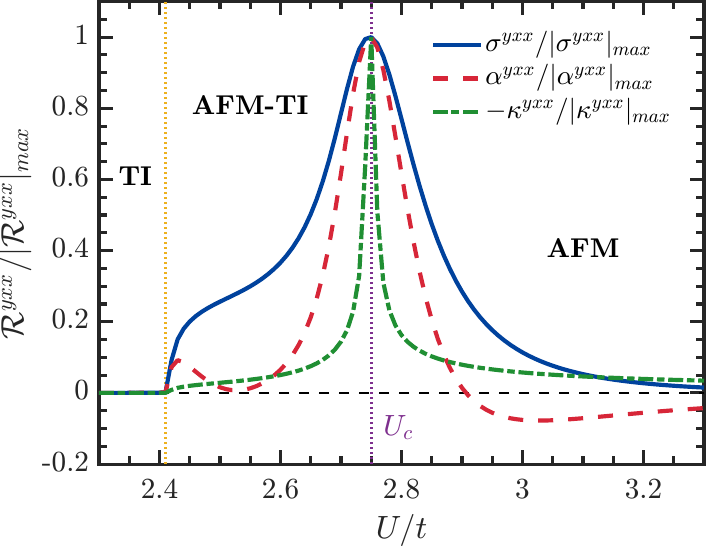}
\caption{\label{fig:transport}
\textbf{Giant nonlinear anomalous responses at the topological critical point.}
Calculated non-zero components of the nonlinear anomalous electrical ($\sigma^{yxx}$), thermoelectric ($\alpha^{yxx}$), and thermal ($\kappa^{yxx}$) conductivities as a function of the correlation $U$. All coefficients are normalized by their respective absolute maximum values, with the sign of $\kappa_{yxx}$ inverted for visual comparison.}
\end{figure}

\CV \textit{Nonlinear anomalous transport responses.-} \CIV
Based on the above analysis, we now evaluate the nonlinear anomalous transport coefficients across the phase transitions [see \hyperref[fig:transport]{Fig.~\ref*{fig:transport}}] for a fixed $\lambda = 0.05t$, utilizing the identical parameters as in \hyperref[fig:phase]{Fig.~\ref*{fig:phase}(a)}.
Within this generalized parameter space, the electron correlations play two roles in governing the transport properties. First, in the weak-interaction TI limit, all coefficients vanish due to the preserved $\mathcal{P}$ symmetry.
As $U$ increases, the system undergoes a symmetry-breaking transition: the transition to the AFM-TI phase induces a spontaneous staggered magnetization that breaks $\mathcal{P}$ and $\mathcal{T}$, thereby activating the intrinsic anomalous nonlinear responses. 
Further increasing $U$ drives the system toward the topological critical point. At the critical boundary separating the AFM-TI and AFM phases ($U=U_c$), the correlation-driven bulk gap closure triggers a divergence in the underlying quantum geometric tensors.
As shown in \hyperref[fig:transport]{Fig.~\ref*{fig:transport}}, this criticality manifests as a sharp peak across all transport coefficients ($\sigma^{yxx}$, $\alpha^{yxx}$, and $\kappa^{yxx}$), marking a strong enhancement of the nonlinear anomalous responses. 
This giant enhancement of the nonlinear anomalous responses may act as a macroscopic signature of the topological phase transition. 
The initial onset of these signals at the magnetic transition, paired with their subsequent rapid decay in the AFM phase, maps out the boundaries of the AFM-TI phase.
This entire evolution demonstrates how electron correlations dynamically dictate both the magnetic order and the band topology.
This correlation-driven giant anomalous transport enhancement validates our proposal of using intrinsic nonlinear anomalous transport as a macroscopic probe. 

\begin{figure}[t]
\centering
\includegraphics[width=\columnwidth]{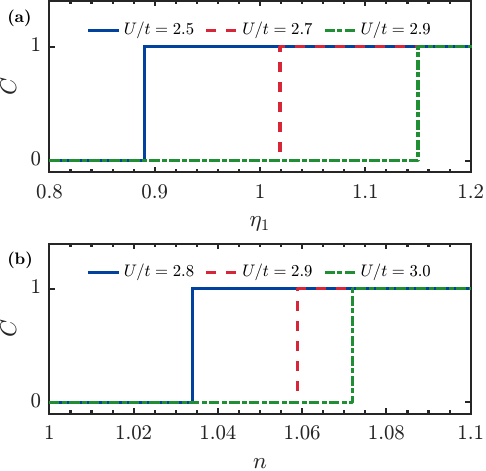}
\caption{\label{fig:tuning}
\textbf{Tunability of the topological phase transitions.}
Evolution of the band Chern number $C$, which demarcates the topological transition of the AFM-TI phase, as a function of \textbf{(a)} the strain parameter $\eta_1$ and \textbf{(b)} the electron filling $n=n_A=n_B$ for various interaction strengths $U$.}
\end{figure}

\CV \textit{Discussion and Conclusion.-} \CIV
Although the interaction strength $U$ is usually not regarded as a tunable parameter, the system can be driven into the $\mathcal{PT}$-symmetric AFM-TI phase by other external parameters for a fixed $U$. Therefore the enhanced topological thermoelectric response can be effectively reached by tuning lab-controlled parameters. This would facilitate future test in experiment and possible realistic applications. 

First, mechanical strain provides an effective tuning parameter. As shown in \hyperref[fig:tuning]{Fig.~\ref*{fig:tuning}(a)}, varying the hopping anisotropy $\eta_1$ drives a topological phase boundary, switching the system into or out of the AFM-TI phase. The critical value $\eta_1 = 1.1$ corresponds to an uniaxial tensile strain of $2\% - 4\%$ across various two-dimensional materials (see \cite{SM}), which is accessible using modern piezoelectric actuators or flexible substrates \cite{Qin2021,Amorim2016,Azizi2025,Conley2013}.

Second, electrostatic gating allows for \textit{in situ} tuning. \hyperref[fig:tuning]{Fig.~\ref*{fig:tuning}(b)} demonstrates that a change in electron filling $n$ is sufficient to trigger the topological transition bounding the AFM-TI phase. The critical filling $n \approx 1.05$ corresponds to an effective carrier density of $\sim 10^{14}\text{ cm}^{-2}$, well within the operational range of established ionic-liquid gating techniques \cite{Ye2012,Bisri2017,Saito2016,Qin2022,Kim2024}.

In summary, we identify a 2D correlation-induced $\mathcal{PT}$-symmetric AFM-TI phase and reveal a topological phase transition between this phase and the AFM phase within an extended Kane–Mele–Hubbard model.
Remarkably, this topological phase transition yields a giant enhancement of the nonlinear ANE, AHE, and ATHE responses.
The nontrivial interplay between electron correlations and topological characteristics gives rise to the prominent anomalous thermoelectric transport behaviors uncovered in this work.
This appealing correlation-driven topological mechanism offers promising prospects for the future practical application of topological thermoelectric devices.

\CV \textit{Acknowledgments.-} \CIV
This work is supported by the National Key R\&D Program of China (Grant No. 2024YFA1409200, No. 2022YFA1402802), CAS Project for Young Scientists in Basic Research Grant No. YSBR-057.  G.S. was supported in part by the Quantum Science and Technology-National Science and Technology Major Project under Grant No. 2024ZD0300500, NSFC Nos. 12534009 and 12447101, the Strategic Priority Research Program of CAS (Grant No. XDB1270000) and the CAS Superconducting Research Project under Grant No. SCZX-0101.

\bibliography{references}

\end{document}